\documentclass{ws-ijmpe}

\begin{document}

\markboth{M. Thoennessen}{2023 Update of the Discoveries of Isotopes}

\catchline{}{}{}{}{}
%%%%%%%%%%%%%%%%%%%%%%%%%%%%%%%%%%%%%%%%%%%%%%%%%%%%%%%%%%%%%%%%%%%%

\title{2023 UPDATE OF THE DISCOVERIES OF NUCLIDES}

\author{\footnotesize M. THOENNESSEN\footnote{This work was supported by Michigan State University}}

\address{Facility for Rare Isotope Beams \\
Michigan State University\\
East Lansing, Michigan 48824, USA\\
thoennessen@frib.msu.edu}

\maketitle

\begin{history}
\received{Day Month Year}
\revised{Day Month Year}
%\accepted{Day Month Year}
%\comby{(xxxxxxxxxx)}
\end{history}

\begin{abstract}
The 2023 update of the discovery of nuclide project is presented when thirteen nuclides were observed for the first time. In addition, a major update and revision of the isotope discovery project is described.
\end{abstract}

\keywords{Discovery of nuclides; discovery of isotopes}

\ccode{PACS numbers: 21.10.-k, 29.87.+g}

%\tableofcontents

\section{Introduction}

This is the eighth update of the isotope discovery project which was originally published in a series of papers in Atomic Data and Nuclear Data Tables from 2009 through 2013 (see for example the first\cite{2009Gin01} and last\cite{2013Fry01} papers). Two summary papers were published in 2012 and 2013 in Nuclear Physics News\cite{2012Tho03} and Reports on Progress in Physics,\cite{2013Tho02} respectively, followed by annual updates in 2014,\cite{2014Tho01} 2015,\cite{2015Tho01} 2016,\cite{2016Tho02} 2017,\cite{2017Tho01} 2018,\cite{2018Tho01} 2019,\cite{2019Tho01} and 2023.\cite{2023Tho01} The last update included the isotopes discovered between 2019 and 2022. In 2016 a description of the discoveries from an historical perspective was published in the book ``The Discovery of Isotopes--A complete Compilation.''\cite{2016Tho01}

\section{New discoveries in 2023}
\label{New2023}

The 13 isotopes discovered in 2023 continue the trend of about 10 discoveries per year over the last five years. This is in contrast to the previous five years (2014--2018) when on average 25 isotopes were discovered. Table \ref{2023Isotopes} lists details of the discoveries including the production method. The table also includes the discovery of $^{116}$La  which was already reported at the end of 2022 but were not included in last year's update. It represented the first isotope discovery in the journal Communications Physics. This increased the number of isotopes discovered in 2022 to twelve. 

\begin{table}[pt]
\tbl{New nuclides reported in 2023. The nuclides are listed with the first author, submission date, and reference of the publication, the laboratory where the experiment was performed, and the production method (PF = projectile fragmentation, FE = fusion evaporation, SB = secondary beams, TR = transfer reactions). \label{2023Isotopes}}
{\begin{tabular}{@{}llrclc@{}} \toprule 
Nuclide(s) & First Author & Subm. Date & Ref. & Laboratory & Type \\ \colrule
 $^{116}$La$^a$ & W. Zhang & 7/5/2022 & \refcite{2022Zha01} & Jyv\"askyl\"a & FE \\
 $^{27}$O, $^{28}$O & Y. Kondo & 10/13/2022 & \refcite{2023Kon01} & RIKEN & SB \\
 $^{9}$N & R. J. Charity & 11/2/2022 & \refcite{2023Cha01} & MSU & SB \\
 $^{241}$U & T. Niwase & 11/21/2022 & \refcite{2023Niw01} & RIKEN & TR \\
 $^{190}$At & H. Kokkonen & 3/20/2023 & \refcite{2023Kok01} & Jyv\"askyl\"a & FE \\
 $^{189}$Lu, $^{191}$Hf, $^{192}$Hf & K. T. Haak & 6/27/2023 & \refcite{2023Haa01} & MSU & PF \\
 $^{156}$W,$^b$ $^{160}$Os$^b$ & H. B. Yang & 7/5/2023 & \refcite{2024Yan01} & Lanzhou & FE \\
 $^{276}$Ds, $^{272}$Hs, $^{268}$Sg & Yu. Ts. Oganessian & 7/23/2023 & \refcite{2023Oga01} & Dubna & FE \\
\botrule
\vspace*{-0.2cm} & & & \\
$^a$ discovered already in 2022\\
$^b$ discovered in 2024, see text
\end{tabular}}
\end{table}

$^{116}$La was discovered by Zhang et al. and reported in the paper ``Observation of the proton emitter $^{116}_{57}$La$^{59}$.''\cite{2022Zha01} The K130 cyclotron at the Accelerator Laboratory of the University of Jyv\"asky\"a in Finland accelerated a $^{64}$Zn beam to 330 MeV. $^{116}$La was produced in the $p5n$ reaction on an isotopically enriched 750$\mu$g $^{58}$Ni target and identified with the vacuum-mode recoil separator MARA where the residues were detected in a double sided silicon strip detector. Evaporated charged-particles were detected with the JYTube (Jyv\"askyl\"a-York Tube) array. ``The extremely neutron deficient isotope $^{116}$La has been observed via its ground-state proton emission ($E_p$ = 718(9) keV, $T_{1/2}$ = 50(22) ms).''

Kondo et al. reported the discovery of $^{27}$O and $^{28}$O in ``First Observation of $^{28}$O.''\cite{2023Kon01} A primary $^{48}$Ca beam accelerated to 345 MeV/A by the RI Beam Factory at RIKEN was used to produce secondary beams of 235 MeV/A $^{29}$F and $^{29}$Ne which then impinged on a 151 mm liquid hydrogen target. $^{28}$O was populated in the one-proton removal reaction from $^{29}$F. $^{27}$O was identified in the subsequent neutron emission from $^{28}$O as well as in the two-proton removal reaction from $^{29}$Ne. Charged reaction fragments were separated and identified with the SAMURAI spectrometer and coincident neutrons were measured with the large-area segmented plastic scintillator walls NeuLAND and NEBULA. In addition, the reaction vertex was determined by the MINOS Time Projection Chamber. ``In the case of $^{27}$O, a decay energy of E$_{0123}$ = 1.09$\pm$0.04(stat)$\pm$0.02(syst) MeV was found,'' and ``In the case of $^{28}$O, a decay energy of E$_{01234}$ = 0.46$^{+0.05}_{-0.04}$(stat)$\pm$0.02(syst) MeV was found, with an upper limit of the width of the resonance of 0.7 MeV (68\% confidence interval).''

In the paper ``Strong evidence for $^9$N and the limits of existence of atomic nuclei'' Charity et al. reported the first observation of $^9$N.\cite{2023Cha01} The Coupled Cyclotron Facility at the National Superconducting Cyclotron Laboratory at Michigan State University was used to produce a secondary 68.5 MeV $^{13}$O beam which impinged on a 1-mm thick $^9$Be target. Charged particles from the reactions were measured with the high resolution array HiRA covering angles between 2.1$^\circ$--12.4$^\circ$, and invariant mass spectra of events containing five protons and one $\alpha$ particle were reconstructed. ``We have found strong evidence for the exotic nuclide $^9$N produced in the fragmentation of a $^{13}$O beam. The invariant-mass spectrum of detected 5$p + \alpha$ events, each containing an $^8$C(g.s.) intermediate state, contains a structure which cannot be explained at the $\sim$5$\sigma$ level by statistical fluctuations of the expected background from other coincident protons liberated in the fragmentation event.''

The first observation of $^{241}$U was reported by Niwase et al. in the paper ``Discovery of New Isotope $^{241}$U and Systematic High-Precision Atomic Mass Measurements of Neutron-Rich Pa-Pu Nuclei Produced via Multinucleon Transfer Reactions.''\cite{2023Niw01} The RIKEN Ring Cyclotron accelerated $^{238}$U ions to 10.75 MeV/nucleon which were then focussed on a rotating 12.5 mg/cm$^2$ enriched $^{198}$Pt target. $^{241}$U was produced in the multinucleon transfer reaction $^{198}$Pt($^{238}$U,$^{241}$U) and identified with the KEK Isotope Separation System (KISS). The mass of $^{241}$U was measured with a multireflection time-of-flight mass spectrograph (MRTOF MS). ``The first identification of $^{241}_{92}$U, produced by an MNT reaction in the $^{238}_{92}$U + $^{198}_{78}$Pt system, was made by mass spectrometry.''

The discovery of $^{190}$At was reported by Kokkonen et al. in ``Properties of the new $\alpha$-decaying isotope $^{190}$At.''\cite{2023Kok01} The K-130 cyclotron at the Accelerator Laboratory of the University of Jyv\"askyl\"a (JYFL) accelerated a $^{84}$Sr beam to 380$-$390 MeV which then impinged on a natural tin target. $^{190}$At was formed in the fusion evaporation reaction $^{109}$Ag($^{84}$Sr,3n). Evaporation residues were separated with the gas-filled recoil separator RITU (Recoil Ion Transport Unit) and identified with the GREAT (Gamma Recoil Electron Alpha Tagging) spectrometer. ``An $\alpha$-particle energy of 7750(20) keV and a half-life of 1.0$^{+1.4}_{-0.4}$ ms were measured.''

Haak et al. discovered $^{189}$Lu, $^{191}$Hf, and $^{192}$Hf in ``Production and discovery of neutron-rich isotopes by fragmentation of $^{198}$Pt.''\cite{2023Haa01} The Coupled Cyclotron Facility at the NSCL on the campus of Michigan State University accelerated a $^{198}$Pt beam to 85 MeV/u which then impinged on a 47 mg/cm$^2$ beryllium target. Fragmentation projects were separated with a combination of the A1900 fragment separator and the S800 analysis beam line and identified in a PIN diode telescope.  ``During the course of the experiment and the measurement of production cross sections, three new isotopes, namely $^{189}$Lu and $^{191,192}$Hf, were discovered. [...] The measured cross sections of these nuclides were found to be 0.037(24), 0.13(5), and 0.061(44) nb, respectively.''

The discovery of $^{156}$W and $^{160}$Os was first published in November 2023 by Briscoe et al. in ``Decay spectroscopy at the two-proton drip line: Radioactivity of the new nuclides $^{160}$Os and $^{156}$W.''\cite{2023Bre01} However, the paper was submitted on 7/11/2023, six days after (7/5/2023) the submission by Yang et al. entitled ``Discovery of New Isotopes $^{160}$Os and $^{156}$W: Revealing Enhanced Stability of the N=82 Shell Closure on the Neutron-Deficient Side.''\cite{2024Yan01} Thus the primary credit for the discovery is given to Yang et al. Isotopically enriched $^{106}$Cd targets were irradiated with a 335 MeV $^{58}$Ni beam accelerated by the Sector Focusing Cyclotron of the Heavy Ion Research Facility at Lanzhou. Evaporation residues from the reaction $^{106}$Cd($^{58}$Ni,4n)$^{160}$Os were separated with the Spectrometer for Heavy Atoms and Nuclear Structure (SHANS) and implanted in three position-sensitive silicon strip detectors. These detectors as well as eight other silicon detectors surrounding the implantation detectors recorded correlated $\alpha$-particles. ``The measured $\alpha$-particle energy and half-life values of $^{160}$Os are 7080(26) keV and 201$^{+58}_{-37}$ $\mu$s, respectively. The half-life of $^{156}$W was determined to be 291$^{+86}_{-61}$ ms.'' Yang et al. acknowledge the work by Briscoe et al. in a note added: ``Recently, a parallel effort to discover $^{160}$Os and $^{156}$W \cite{2023Bre01} was published. We note that our results are in agreement with the reported data within the experimental accuracy.''

%The K130 cyclotron at the Accelerator Laboratory of the University of Jyv\"askyl\"a provided a 310 MeV $^{58}$Ni beam for the fusion evaporation reaction $^{106}$Cd($^{58}$Ni,4n)$^{160}$Os. Residues were implanted in a double-sided silicon strip detector located at the focal plane of the recoil mass separator MARA. Protons and $\alpha$-particles emitted from the target were detected with the charged particle detector array JYTube. ``The $\alpha$ decays of the ground state of $^{160}$Os ($E_\alpha$ = 7092(15) keV, $t_{1/2}$ = 97$^{+97}_{-32} \mu$s) and its isomeric state ($E_\alpha$ = 8890(10) keV, $t_{1/2} = 41^{+15}_{-9} \mu$s) were measured [...] The half-life of $^{156}$W was determined to be 153$^{+64}_{-39}$ ms from the time differences between $^{160}$Os $\alpha$ decays and subsequent $^{156}$Ta proton decays or $^{156}$Hf isomer $\alpha$ decays, taking into account the half-lives of the intermediate $\beta$-decaying states.''

$^{276}$Ds, $^{272}$Hs, and $^{268}$Sg were discovered by Oganessian et al. in ``New isotope $^{276}$Ds and its decay products $^{272}$Hs and $^{268}$Sg from the $^{232}$Th + $^{48}$Ca reaction.''\cite{2023Oga01} The DC280 cyclotron at the SHE Factory at JINR in Dubna accelerated a $^{48}$Ca beam to 230$-$250 MeV impinging on $^{232}$Th targets. Evaporation residues from the reaction $^{232}$Th($^{48}$Ca,4n) were separated and identified with the gas-filled separator DGFRS-2 which also detected subsequent $\alpha$ decays. ``Three new nuclides were synthesized for the first time: a spontaneously fissioning (SF) $^{268}$Sg with the half-life T$_{SF}$ = 13$^{+17}_{-4}$ s, an $\alpha$ decaying $^{272}$Hs with T$_\alpha$ = 0.16$^{+0.19}_{-0.06}$ s, E$_\alpha$ = 9.63$\pm0.02$ MeV, and $^{276}$Ds with T$_{1/2}$ = 0.15$^{+0.10}_{-0.04}$ ms, E$_\alpha$ = 10.75$\pm$0.03 MeV, and an SF branch of 57\%.''

\section{Reassignments in 2023}
\label{Reassign}

In the past, the discovery of a few isotopes could not be assigned to one unique article or laboratory. For ten isotopes two papers reporting the discoveries were submitted on the same date and most of them were published in the same issue of a journal, some back to back. In addition, the research for the discovery of five isotopes was performed at two different laboratories and reported in the same article. In order to avoid giving partial or multiple credit for these isotopes the ambiguities were resolved. For the ten simulateously reported discoveries, preference is now given to the paper which is published first (see table \ref{double}). $^{13}$C is a special case where the authors submitted the same article at the same time to Physical Review\cite{1929Kin01} and Nature.\cite{1929Kin02}

\begin{table}[pt]
\tbl{Simultaneously submitted discoveries; primary credit is now given to the paper listed first. \label{double}}
{\begin{tabular}{@{}rllllll@{}} \toprule 
Isotope & First Author & Ref. & Laboratory & Country & Year & Published \\  \colrule
$^{13}$C & A.S. King & \refcite{1929Kin01} & Carnegie Inst. & USA & 1929 & 7/15/1929\\
 & A.S. King & \refcite{1929Kin02} & Carnegie Inst. & USA & 1929  & 7/27/1929\\
$^{37}$Ca & J.C. Hardy & \refcite{1964Har01} & McGill & Canada & 1964 & page 764\\
 & P.L. Reeder & \refcite{1964Ree01} & Brookhaven & USA & 1964 & next article\\
$^{50}$V & D.C. Hess Jr. & \refcite{1949Hes01} & Argonne & USA & 1949 & page 1717\\
 & W.T. Leland & \refcite{1949Lel01} & Minnesota & USA & 1949 & page 1722\\
$^{145}$Dy & G.D. Alkhazov & \refcite{1982Alk01} & Leningrad & Russia & 1982 & June 1982 \\
 & E. Nolte & \refcite{1982Nol02} & Munich & Germany & 1982 & Sept. 1982\\
$^{165}$Dy & J. K. Marsh & \refcite{1935Mar01} & Oxford & UK & 1935 & page 102\\
 & G. Hevesy & \refcite{1935Hev01} & Copenhagen & Denmark & 1935 & next article\\
$^{245}$Pu & C.I. Browne & \refcite{1955Bro01} & Los Alamos & USA & 1955 & page 254\\
 & P.R. Fields & \refcite{1955Fie01} & Argonne & USA & 1955 & next article\\
$^{245}$Am & C.I. Browne & \refcite{1955Bro01} & Los Alamos & USA & 1955 & page 254\\
 & P.R. Fields & \refcite{1955Fie01} & Argonne & USA & 1955 & next article\\
$^{242}$Cf & T. Sikkeland & \refcite{1967Sik01} & Berkeley & USA & 1967 & page 331\\
 & P.R. Fields & \refcite{1967Fie01} & Argonne & USA & 1967 & page 340\\
$^{243}$Cf & T. Sikkeland & \refcite{1967Sik02} & Berkeley & USA & 1967 & page 333\\
 & P.R. Fields & \refcite{1967Fie01} & Argonne & USA & 1967 & page 340\\
$^{254}$No & E.D. Donets & \refcite{1966Don02} & Dubna & Russia & 1966 & page 257\\
 & B.A. Zager & \refcite{1966Zag01} & Dubna & Russia & 1966 & next article\\
\botrule
\end{tabular}}
\end{table}

$^{93}$Br and $^{94}$Br were discovered during experiments performed with the mass separators OSTIS and ISOLDE at Grenoble and CERN, respectively. Credit for the discovery is assigned to CERN as the ISOLDE collaboration is listed as a co-author on the publication whereas there were no co-authors from Grenoble.\cite{1988Kra01}

The primary credit for the discovery of $^{105}$Nb is given to the Institut f\"ur Kernphysik at the Kernforschungsanlage J\"ulich because most of the data presented were taken with the JOSEF fission product separator at the DIDO reactor located at J\"ulich, Germany. Data were also recorded with the LOHENGRIN separator at the high flux reactor of the Institut Laue-Langevin in Grenoble, France.\cite{1984Shi01}

The discovery of $^{181}$Os was reported by Hofstetter and Daly in 1966.\cite{1966Hof01} They performed experiments at the Argonne 60-in. cyclotron and the Oak Ridge 88-in cyclotron (ORIC).  The high-quality $\gamma$-ray spectra to identify $^{181}$Os were measured with a 65-MeV $^4$He beam which must have been delivered by ORIC as the Argonne cyclotron did not offer that beam.\cite{1960Ram01}

In 1954, Naumann from Princeton University identified $^{188}$Pt in experiments performed at the Nevis and Harvard University synchrocyclotrons.\cite{1954Nau01} The Columbia University Nevis synchrocyclotron is credited with the primary discovery because only these measurements were mentioned in an earlier conference abstract.\cite{1958Fis01}

Finally, $^{169}$Yb was discovered at the Institut f\"ur Physik am Kaiser Wilhelm-Institut f\"ur Medizinische Forschung in Heidelberg and not the Kaiser Wilhelm-Institut f\"ur Chemie in Berlin-Dahlem.\cite{1946Bot01}

\begin{table}[t]
\tbl{Nuclides only reported in proceedings or internal reports until the end of 2023. The nuclide, first author, reference and year of proceeding or report are listed. \label{reports}}
{\begin{tabular}{@{}llrr@{}} \toprule
\parbox[t]{4.4cm}{\raggedright Nuclide(s) } & \parbox[t]{2.3cm}{\raggedright First Author} & Ref. & Year \\ \colrule
$^{21}$C	&	 S. Leblond 	&	\refcite{2015Leb01},\refcite{2015Leb02}	&	2015	 \\ 
			&	 N. A. Orr 	&	\refcite{2016Orr01}	&	2016	 \\ 
$^{24}$N, $^{25}$N		&	 Q. Deshayes 	&	\refcite{2018Des01}	&	2018	 \\ 
$^{45}$Si, $^{46}$Si		&	 H. Suzuki 	&	\refcite{2021Suz01}	&	2021	 \\ 
$^{79}$Co, $^{84}$Cu, $^{86}$Zn, $^{93}$As$^a$ & Y. Shimizu & \refcite{2018Shi02} & 2018 \\
$^{93}$As,$^a$ $^{96}$Se, $^{99}$Br	&	 Y. Shimizu 	&	\refcite{2015Shi01}	&	2015	 \\ 
$^{98}$Sn	&	  I. Celikovic 	&	\refcite{2013Cel01}	&	2013	 \\ 
$^{126}$Nd, $^{136}$Gd, $^{138}$Tb, $^{143}$Ho,$^b$ $^{153}$Hf	&	 G. A. Souliotis 	&	\refcite{2000Sou01}	&	2000	 \\
$^{143}$Ho$^b$	&	 D. Seweryniak	&	\refcite{2003Sew02}	&	2003	 \\
$^{143}$Er, $^{144}$Tm	&	 R. Grzywacz 	&	\refcite{2005Grz01}	&	2005	 \\
	& K. Rykaczewski & \refcite{2005Ryk01} & 2005 \\
	& C. R. Bingham & \refcite{2005Bin01} & 2005 \\
$^{230}$At, $^{232}$Rn	&	 J. Benlliure 	&	\refcite{2010Ben02},\refcite{2015Ben01}	&	2010/15	 \\
$^{252}$Bk, $^{253}$Bk	&	 S. A. Kreek 	&	\refcite{1992Kre01}	&	1992	 \\
$^{262}$No 	&	 R. W. Lougheed 	&	\refcite{1988Lou01},\refcite{1989Lou01}	&	1988/89	 \\
	&	 E. K. Hulet 	&	\refcite{1989Hul01}	&	1989	 \\
$^{261}$Lr, $^{262}$Lr	&	 R. W. Lougheed 	&	\refcite{1987Lou01}	&	1987	 \\
	&	 E. K. Hulet 	&	\refcite{1989Hul01}	&	1989	 \\
	&	 R. A. Henderson 	&	\refcite{1991Hen01}	&	1991	 \\
$^{255}$Db	&	 G. N. Flerov	&	\refcite{1976Fle01}	&	1976	 \\
		& A.-P. Lepp\"anen	& \refcite{2005Lep01} & 2005 \\
$^{275}$Ds& A. Karpov &	\refcite{2023Kar01}	&	2023	 \\
\botrule
\vspace*{-0.2cm} & & & \\
$^a$ published in refs. \refcite{2015Shi01}  and  \refcite{2018Shi02} \\
$^b$ published in refs. \refcite{2000Sou01} and \refcite{2003Sew02} \\
\end{tabular}}
\end{table}

\section{Discoveries not yet published in refereed journals}

Table \ref{reports} lists the isotopes which so far still have only been presented in conference proceedings or internal reports. This year isotopes presented in the 2015 RIKEN Accelerator Progress Report which were only shown in particle identification plots were removed from the list as they were not explicitely mentioned as identified isotopes. This affects ten neutron-rich isotopes for elements between barium and thulium\cite{2015Fuk01} and  
seven isotopes for elements between zinc and bromine.\cite{2015Shi01} Nevertheless, the discovery of these isotopes will hopefully be published soon.

It is interesting to note that the discovery of $^{275}$Ds was first announced in a press release from the Joint Institute for Nuclear Research (JINR) in Dubna\cite{2023SHE01} on March 19, 2023. A couple of months later the discovery was presented at the ISPUN23 conference in Phu Quoc Island, Vietnam.\cite{2023Kar01} The proceeding of this conference has not been published yet.

\section{Status at the end of 2023 -- Summary}

\begin{figure}[pt] 
\centerline{\psfig{file=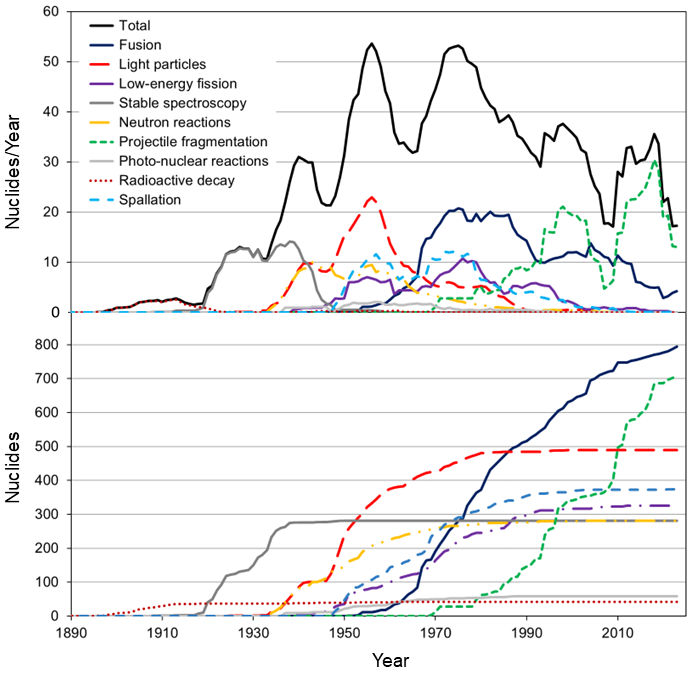,width=11.7cm}}
\caption{Discovery of nuclides as a function of year. The top panel shows the 10-year running average of the number of nuclides discovered per year while the bottom panel shows the cumulative number. The different colored lines correspond to the different methods used to produce the nuclides as shown in the bottom panel. The top panel also shows the total number of nuclides (black line). The figure was adapted from Ref. \protect\refcite{2018Tho01} to include the data from 2023. \label{f:timeline} }
\vspace*{-0.1cm}
\end{figure}

The 13 new discoveries in 2023 increased the total number of observed isotopes to 3352. They were reported by 947 different first authors in 1577 papers and a total of 3973 different coauthors. These numbers for authors, coauthors, and papers are not consistent with the previous update because of the reassignments described in Section \ref{Reassign}. Further statistics can be found on the new discovery project website\cite{2024DIP01} which has been significantly redesigned and which now includes the abstracts describing the discovery details of all individual isotopes. The abstracts can be searched by element, author, lab, country, journal, year, or reference. A clickable chart of nuclides linking to the abstracts is also available at the National Nuclear Data Center.\cite{2024NDC01}

Figure \ref{f:timeline} shows the current status of the evolution of the nuclide discoveries for the main means of production as labeled in the figure. The figure was adapted  from the review of isotopes discovered until 2017\cite{2018Tho01} and was extended to include all isotopes discovered until the end of 2023. The top part of the figure shows the ten-year average of the number of nuclides discovered per year while the bottom panel shows the integral number of nuclides discovered. The top figure includes the total of nuclides while the bottom figure only shows the contributions of the individual production methods.

98\% of the isotopes discovered during the last twenty years were produced by either projectile fragmentation reactions (71\%) or fusion-evaporation reactions (27\%). The fraction of projectile fragmentation to fusion evaporation has increased from about 55\% to 90\% during this time. As a reminder, projectile fragmentation also includes in-flight projectile fission, secondary beams and deep inelastic reactions and fusion-evaporation reactions include heavy-ion transfer reactions.

Only ten isotopes were first observed by other methods. Eight isotopes were discovered by low-energy fission where most recently $^{155}$Pr was identified twelve years ago with the Canadian Penning Trap at Argonne National Laboratory.\cite{2012Van01} The most recent isotope produced in a spallation reaction was $^{150}$Yb in 2021 at the Isotope Separator and Accelerator (ISAC) facility at TRIUMF in Canada.\cite{2021Bec01}

It is also noteworthy that the ten-year running average rate was dominated by Japan and RIKEN over the last two years when they discovered about 66\% of all isotopes. No other country or laboratory has discovered such a large fraction since about 90\% of all isotopes were discovered in the U.S. in the 1950s and in Cambridge in the 1920/30s. No other country or lab discovered more than 10\%.

%\section*{Acknowledgements}

%\vfill
%\newpage

\bibliographystyle{ws-ijmpe}
%\bibliography{test2}
\bibliography{references}

\begin{thebibliography}{10}

\bibitem{2009Gin01}
{J. Q. Ginepro, J. Snyder, and M. Thoennessen}, {\em At. Data Nucl. Data
  Tables} {\bf 95}  (2009)   805.

\bibitem{2013Fry01}
{C. Fry and M. Thoennessen}, {\em At. Data Nucl. Data Tables} {\bf 99}  (2013)
   520.

\bibitem{2012Tho03}
{M. Thoennessen}, {\em Nucl. Phys. News} {\bf 22(3)}  (2012)  ~19.

\bibitem{2013Tho02}
{M. Thoennessen}, {\em Rep. Prog. Phys.} {\bf 76}  (2013)   056301.

\bibitem{2014Tho01}
{M. Thoennessen}, {\em Int. J. Mod. Phys. E} {\bf 23}  (2014)   1430002.

\bibitem{2015Tho01}
{M. Thoennessen}, {\em Int. J. Mod. Phys. E} {\bf 24}  (2015)   1530002.

\bibitem{2016Tho02}
{M. Thoennessen}, {\em Int. J. Mod. Phys. E} {\bf 25}  (2016)   1630004.

\bibitem{2017Tho01}
{M. Thoennessen}, {\em Int. J. Mod. Phys. E} {\bf 26}  (2017)   1730003.

\bibitem{2018Tho01}
{M. Thoennessen}, {\em Int. J. Mod. Phys. E} {\bf 27}  (2018)   1830002.

\bibitem{2019Tho01}
{M. Thoennessen}, {\em Int. J. Mod. Phys. E} {\bf 28}  (2019)   1930002.

\bibitem{2023Tho01}
{M. Thoennessen}, {\em Int. J. Mod. Phys. E} {\bf 32}  (2023)   2330001.

\bibitem{2016Tho01}
{{M. Thoennessen, {\it The Discovery of Isotopes, A Complete Compilation},
  Springer International Publishing 2016, doi:10.1007/978-3-319-31763-2} }.

\bibitem{2022Zha01}
{W. Zhang {\it et al.}}, {\em Commun. Phys.} {\bf 5}  (2022)   285.

\bibitem{2023Kon01}
{Y. Kondo {\it et al.}}, {\em Nature} {\bf 620}  (2023)   965.

\bibitem{2023Cha01}
{R. J. Charity {\it et al.}}, {\em Phys. Rev. Lett.} {\bf 131}  (2023)
  172501.

\bibitem{2023Niw01}
{T. Niwase {\it et al.}}, {\em Phys. Rev. Lett.} {\bf 130}  (2023)   132502.

\bibitem{2023Kok01}
{H. Kokkonen {\it et al.}}, {\em Phys. Rev. C} {\bf 127}  (2023)   064312.

\bibitem{2023Haa01}
{K. Haak {\it et al.}}, {\em Phys. Rev. C} {\bf 108}  (2023)   034608.

\bibitem{2024Yan01}
{H. B. Yang {\it et al.}}, {\em Phys. Rev. Lett.} {\bf 132}  (2024)   072502.

\bibitem{2023Oga01}
{Yu. Ts. Oganessian {\it et al.}}, {\em Phys. Rev. C} {\bf 108}  (2023)
  024611.

\bibitem{2023Bre01}
{A. D. Briscoe {\it et al.}}, {\em Phys. Lett. B} {\bf 847}  (2023)   138310.

\bibitem{1929Kin01}
{A. S. King and R. T. Birge}, {\em Phys. Rev.} {\bf 34}  (1929)   376.

\bibitem{1929Kin02}
{A. S. King and R. T. Birge}, {\em Nature} {\bf 124}  (1929)   127.

\bibitem{1964Har01}
{J. C. Hardy and R. I. Verrall}, {\em Phys. Rev. Lett.} {\bf 13}  (1964)   764.

\bibitem{1964Ree01}
{P. L. Reeder, A. M. Poskanzer, and R. A. Esterlund}, {\em Phys. Rev. Lett.}
  {\bf 13}  (1964)   767.

\bibitem{1949Hes01}
{D. C. {Hess Jr.} and M. G. Inghram}, {\em Phys. Rev.} {\bf 76}  (1949)   1717.

\bibitem{1949Lel01}
{W. T. Leland}, {\em Phys. Rev.} {\bf 76}  (1949)   1722.

\bibitem{1982Alk01}
{G. D. Alkhazov {\it et al.}}, {\em Z. Phys. A} {\bf 305}  (1982)   185.

\bibitem{1982Nol02}
{E. Nolte {\it et al.}}, {\em Z. Phys. A} {\bf 306}  (1982)   223.

\bibitem{1935Mar01}
{J. K. Marsh and S. Sugden}, {\em Nature} {\bf 136}  (1935)   102.

\bibitem{1935Hev01}
{G. Hevesy and H. Levi}, {\em Nature} {\bf 136}  (1935)   103.

\bibitem{1955Bro01}
{C. I. Browne {\it et al.}}, {\em J. Inorg. Nucl. Chem.} {\bf 1}  (1955)   254.

\bibitem{1955Fie01}
{P. R. Fields {\it et al.}}, {\em J. Inorg. Nucl. Chem.} {\bf 1}  (1955)   262.

\bibitem{1967Sik01}
{T. Sikkeland and A. Ghiorso}, {\em Phys. Lett. B} {\bf 24}  (1967)   331.

\bibitem{1967Fie01}
{P. R. Fields {\it et al.}}, {\em Phys. Lett. B} {\bf 24}  (1967)   340.

\bibitem{1967Sik02}
{T. Sikkeland {\it et al.}}, {\em Phys. Lett. B} {\bf 24}  (1967)   333.

\bibitem{1966Don02}
{E. D. Donets, V. A. Shchegolev, and V. A. Ermakov}, {\em Sov. At. Energy} {\bf
  20}  (1966)   257.

\bibitem{1966Zag01}
{B. A. Zager {\it et al.}}, {\em Sov. At. Energy} {\bf 20}  (1966)   264.

\bibitem{1988Kra01}
{K.-L. Kratz {\it et al.}}, {\em Z. Phys. A} {\bf 330}  (1988)   229.

\bibitem{1984Shi01}
{K. Shizuma {\it et al.}}, {\em Z. Phys. A} {\bf 315}  (1984)  ~65.

\bibitem{1966Hof01}
{K. J. Hofstetter and P. J. Daly}, {\em Phys. Rev.} {\bf 152}  (1966)   1050.

\bibitem{1960Ram01}
{W. J. Ramler, J. L. Yntema and M. Oselka}, {\em Nucl. Instrum. Meth.} {\bf 8}
  (1960)   217.

\bibitem{1954Nau01}
{R. A. Naumann}, {\em Phys. Rev.} {\bf 96}  (1954)  ~90.

\bibitem{1958Fis01}
{P. S. Fisher and R. A. Naumann}, {\em Bull. Am. Phys. Soc.} {\bf 3}  (1958)
  209.

\bibitem{1946Bot01}
{W. Bothe}, {\em Z. Naturforsch.} {\bf 1}  (1946)   173.

\bibitem{2015Leb01}
{{S. Leblond {\it et al.}, RIKEN Accel. Prog. Rep. 2014 {\bf 48} (2015) 42} }.

\bibitem{2015Leb02}
{{S. Leblond, PhD Thesis, Universit\'e de Caen (2015);
  https://tel.archives-ouvertes.fr/tel-01289381} }.

\bibitem{2016Orr01}
{N. A. Orr}, {\em EPJ Web Conf.} {\bf 113}  (2016)   06011.

\bibitem{2018Des01}
{{Q. Deshayes, PhD Thesis, Universit\'e de Caen (2018);
  https://tel.archives-ouvertes.fr/tel-01696611}}.

\bibitem{2021Suz01}
{{H. Suzuki {\it et al.}, RIKEN Accel. Prog. Rep. 2020 {\bf 54} (2021) 2} }.

\bibitem{2018Shi02}
{{Y. Shimizu {\it et al.}, RIKEN Accel. Prog. Rep. 2017 {\bf 51} (2018) 84}}.

\bibitem{2015Shi01}
{{Y. Shimizu {\it et al.}, RIKEN Accel. Prog. Rep. 2014 {\bf 48} (2015) 71} }.

\bibitem{2013Cel01}
{{I. Celikovic, PhD Thesis, Universit\'e de Caen (2013);
  https://theses.hal.science/tel-00981493} }.

\bibitem{2000Sou01}
{G. A. Souliotis}, {\em Physica Scripta} {\bf T88}  (2000)   153.

\bibitem{2003Sew02}
{{D. Seweryniak {\it et al.}, ANL-03/23 (Physics Division Ann. Rept., 2002),
  p.31 (2003)}}.

\bibitem{2005Grz01}
{R. Grzywacz {\it et al.}}, {\em Eur. Phys. J. A} {\bf 25}  (2005)   s145.

\bibitem{2005Ryk01}
{K. P. Rykaczewski {\it et al.}}, {\em AIP Conf. Proc.} {\bf 764}  (2005)
  223.

\bibitem{2005Bin01}
{C. R. Bingham {\it et al.}}, {\em Nucl. Instrum. Meth. B} {\bf 241}  (2005)
  185.

\bibitem{2010Ben02}
{{J. Benlliure {\it et al.}, Pos(NIC XI) 084 (2010)}}.

\bibitem{2015Ben01}
{J. Benlliure}, {\em EPJ Web Conf.} {\bf 88}  (2015)   00028.

\bibitem{1992Kre01}
{{S. A. Kreek {\it et al.}, LBL Nuclear Science Division Annual Report for
  1991, LBL--31855, p. 57 (1992)}}.

\bibitem{1988Lou01}
{{R. W. Lougheed {\it et al.}, Lawrence Livermore Nat. Lab. Rept. UCAR
  10062-88, 135 (1988)}}.

\bibitem{1989Lou01}
{{R. W. Lougheed {\it et al.},``50 Years With Nuclear Fission'' Conference,
  Vol. II, p. 694, Washington, DC, April 25–28, 1989, publ. by Amer. Nucl.
  Soc. Inc., LaGrange, IL 60525 (1989)}}.

\bibitem{1989Hul01}
{{E. K. Hulet, UCRL-100763 (1989)}}.

\bibitem{1987Lou01}
{{R. W. Lougheed {\it et al.}, Lawrence Livermore Nat. Lab. Rept. UCAR
  10062-87, 4-2 (1987)}}.

\bibitem{1991Hen01}
{{R. A. Henderson {\it et al.}, LBL-30798, p. 65 (1991)}}.

\bibitem{1976Fle01}
{{G.N. Flerov, Proc. of the Third Int. Conf. on Nuclei Far from Stability,
  Cargese, Corsica, France 1976, CERN-76-13, p. 542 (1976)}}.

\bibitem{2005Lep01}
{{A.-P. Lepp\"anen, Department of Physics, University of Jyv\"askyl\"a Research
  Report, No. 5/2005} }.

\bibitem{2023Kar01}
{{A. Karpov, ISPUN23, May 4-8, 2023, Phu Quoc Island, Vietnam,
  https://ispun23.vn/public/presents/Karpov\_SHE.pdf}}.

\bibitem{2015Fuk01}
{{N. Fukuda {\it et al.}, RIKEN Accel. Prog. Rep. 2014 {\bf 48} (2015) 72} }.

\bibitem{2023SHE01}
{{JINR Press Release: New darmstadtium isotope discovered at Superheavy Element
  Factory, March 17, 2023,
  http://www.jinr.ru/posts/new-darmstadtium-isotope-discovered-at-superheavy-element-factory/}}.

\bibitem{2024DIP01}
{{https://frib.msu.edu/public/nuclides/}}.

\bibitem{2024NDC01}
{{https://www.nndc.bnl.gov/nudat3/ NuDat 3 is actively maintained and updated
  by the National Nuclear Data Center (NNDC) at Brookhaven National
  Laboratory}}.

\bibitem{2012Van01}
{J. {Van Schelt} {\it et al.}}, {\em Phys. Rev. C} {\bf 85}  (2012)   045805.

\bibitem{2021Bec01}
{S. Beck {\it et al.}}, {\em Phys. Rev. Lett.} {\bf 127}  (2021)   112501.

\end{thebibliography}

\end{document}